# Weyl-Heisenberg Transform Capabilities in JPEG Compression Standard

V.M. Asiryan
*Department of Computer Science*
*National University of Science and Technology "MISiS"*
Moscow, Russia
dmc5mod@yandex.ru

V.P. Volchkov
*General Communication Theory chair*
*Moscow Technical University of Communications and Informatics*
Moscow, Russia
volchkovvalery@mail.ru

N.V. Papulovskaya
*Department of Information Technology and Control Systems*
*Ural Federal University named after first President of Russia B. N. Yeltsin*
Ekaterinburg, Russia
n.v.papulovskaia@urfu.ru

*Abstract*—This paper is devoted to the development and research of a new compression technology based on Weyl-Heisenberg bases (WH-technology) for modifying the JPEG compression standard and improving its characteristics. For this purpose, the paper analyzes the main stages of the JPEG compression algorithm, notes its key features and problems that limit further enhancement of its efficiency. To overcome these limitations, it is proposed to use the real version of the two-dimensional discrete orthogonal Weyl-Heisenberg transform (DWHT) instead of the discrete cosine transform (DCT) at the stage of transformation coding. This transformation, unlike DCT, initially has a block structure and is built on the basis of the Weyl-Heisenberg optimal signal basis, the functions of which are orthogonal and well localized both in the frequency and time domains. This feature of DWHT allows for more efficient decorrelation and compression of element values in each block of the image after transformation coding. As a result, it is possible to obtain more efficient selection and screening of insignificant elements at the subsequent stages of quantization and information coding.

Based on DWHT, a new version of the JPEG compression algorithm was developed, and convenient criteria for evaluating the compression efficiency and metrics of quality losses were proposed. The results of an experimental study are presented, confirming the higher compression efficiency of the proposed algorithm in comparison with the JPEG compression standard.

*Keywords—image compression, JPEG codec, discrete cosine transform, Weyl-Heisenberg transform, downsampling, quantization*

## I. Introduction

With the advent of the possibility of displaying graphic information by personal computers, a concomitant problem has arisen associated with the efficient storage and transmission of raster images. To solve this problem, a committee of experts from around the world was formed in 1986, called the Joint Photographic Experts Group (JPEG). It was this expert group that later in 1992 developed the well-known JPEG raster image compression standard [1]. Today JPEG is the most common format for encoding, sending and storing raster images, without which it is impossible to imagine modern information systems, cameras, personal computers and mobile devices. Several billion JPEG images have been created every day since 2015.

In the original specification, the JPEG compression format is based on earlier researches, works, and patents. The discrete cosine transform (DCT) was announced in 1972 by Nasir Ahmed as an image compression method. Later, N. Ahmed, together with other researchers, developed a practical algorithm for the discrete cosine transform [2], which attracted the attention of other scientists [3-5], whose overall contribution became fundamental for the JPEG compression standard.

Today, the results of modern scientific research in the field of frequency and time-frequency characteristics of signals provide an opportunity to develop a more efficient compression algorithms. As noted in [6], the main task is to synthesize a universal basis. Such a basis should be able to functionally separate the signal in the time-frequency domain and some fragments. Then we have the opportunity to analyze the spectral features of the signal.

One of the possible solutions to this problem is the short-time Fourier transform, which allows one to obtain a characteristic of the signal frequency distribution over time. But the main problem when using the short Fourier transform is associated with the Heisenberg uncertainty principle, which operates with respect to the parameters of the time and frequency of the signal. This principle is based on the fact that it is impossible to say exactly at what point in time a certain frequency is present in the signal, we can only talk about a time interval or frequency range. In addition, the short-time Fourier transform is redundant because it is a complex transform and, as a result, is not suitable as an image compression tool.

In turn, the use of wavelet transforms (Wavelet transform), developed as a tool that solves the Heisenberg uncertainty problem for obtaining the time-frequency characteristics of the signal and which are widely used in image compression problems – JPEG 2000 [7], is not always the best tool. In addition, a common disadvantage of wavelets is the asymmetry of the forming functions.

It is for these reasons that one of the most promising methods for obtaining the frequency-time characteristics of a signal is the use of generalized Weyl-Heisenberg bases obtained by uniform shifts in time and frequency of one or several functions [8]. In early works it was shown that the Weyl-Heisenberg basis, built on the basis of the Gaussian function, is of particular interest for research. Today, modern approaches and algorithms aimed at synthesizing the optimal large Weyl-Heisenberg basis make it possible to apply it to large signals, in particular images [9].

As noted earlier, this article focuses on the implementation of Weyl-Heisenberg bases compression technology to modify the JPEG compression standard. Although the JPEG standard is not relatively new, but it still has efficient compression characteristics. The ideas of the JPEG standard formed the basis for later compression algorithms such as JPEG2000, HEIF, and others. To better understand the proposed solution for modifying the JPEG standard, the main steps of the JPEG compression algorithm are described in more detail below, including color conversion, downsampling, block discrete cosine transform, quantization and entropy coding. Also, the features of the JPEG standard are studied, and a new approach to compression of raster images is proposed, based on the use of the discrete orthogonal Weyl-Heisenberg transform [10]. For this, within the framework of this study, a real Weyl-Heisenberg transform is constructed, recommendations are formulated for choosing the basis parameters, appropriate criteria for evaluating the compression efficiency, metrics of quality loss are introduced, and an objective comparison of the proposed approach to image compression with the JPEG standard is made.

## II. JPEG Codec

As noted earlier, the JPEG compression standard is based on a discrete cosine transform (DCT). This transformation is orthogonal, and therefore easily reversible. By definition, any discrete orthogonal transform is linear and has a matrix representation. In early work [10] we define one-dimensional and two-dimensional discrete orthogonal transforms and show that the most important property that transformation matrix mast have is unitarity. The property of unitarity is described by the expression

$$\mathbf{U}^{*}\mathbf{U} = \mathbf{U}\mathbf{U}^{*} = \mathbf{I}, \qquad (1)$$

where $\mathbf{U}$ – square transformation matrix with discrete orthonormal basis functions (vectors) along the columns, $\mathbf{I}$ – identity matrix.

The matrix of the discrete cosine transform of dimension ($N \times N$) is determined according to the following expression

$$\mathbf{U}_{DCT}(i, j) = \begin{cases} \sqrt{\dfrac{2}{N}} \cos\left(\dfrac{\pi(2j+1)i}{2N}\right), & \text{if } i \neq 0 \\ \sqrt{\dfrac{1}{N}}, & \text{if } i = 0. \end{cases} \qquad (2)$$

$$i = 0, ..., N-1, \; j = 0, ..., N-1.$$

and satisfies property (1).

In fact, DCT represents any finite sequence of data (samples) as a linear combination of discrete basis cosine functions (2) with different frequencies. It should be noted that the discrete cosine transform is closely related to the discrete Fourier transform (DFT) and this relationship is a homomorphism. Therefore, in practice, as a rule, DCT is calculated using fast Fourier transform [11]. However, in this research, this nuance is not fundamental, therefore, the DCT will be presented in matrix form (2).

Before applying the block DCT, the original raster image in terms of red, green and blue (**RGB**) channels is converted to the $\mathbf{YC}_B\mathbf{C}_R$ color space [1]. In this color space, the $\mathbf{Y}$ component represents the luminance of the pixel, and the $\mathbf{C}_B$ and $\mathbf{C}_R$ are the blue-difference and red-difference chroma components. The conversion from **RGB** to $\mathbf{YC}_B\mathbf{C}_R$ color space provides a high compression ratio without significantly affecting perceived image quality. This is explained by the fact that information about brightness is more important for the final quality of image perception, while chromaticity can be partially neglected. For this, color subsampling is used – a decrease in the sampling frequency by decimating the samples of the $\mathbf{C}_B$ and $\mathbf{C}_R$ color components. In the JPEG compression

«*» – conjugate transpose sign.

standard, color downsampling is applied in a 4:2:0 ratio, that is, decimation occurs through one line and one column [1]. This procedure leads to a 2-fold reduction in the dimensionality of the $\mathbf{C}_B$ and $\mathbf{C}_R$ color components both horizontally and vertically. Already at this stage, the original image can be reduced by 1.5 times.

In this article, to simplify the modeling and interpretation of the results, all experiments are carried out with monochrome raster images consisting of one channel – the luminance component $\mathbf{Y}$. Therefore, color conversion and subsampling are not used. However, similar results can be obtained for color images in $\mathbf{YC}_B\mathbf{C}_R$ terms.

According to the JPEG standard [1], the discrete cosine transform is applied not to the entire image as a whole, but separately to each 8×8 image block. The original image is divided into a certain number of blocks, each of which is independently processed. The choice of this block size is due to several reasons. First, an increase in the size of blocks does not lead to a strong increase in compression rates, while the computational complexity of the algorithm increases. Second, there is a high probability of the presence of a large number of sharp boundaries within one block, which can lead to the Gibbs effect.

The forward and backward block discrete orthogonal transforms are written as

$$\mathbf{Z}^{(k)} = \mathbf{U}^{*}\mathbf{Y}^{(k)}\mathbf{U}, \qquad (3)$$

$$\tilde{\mathbf{Y}}^{(k)} = \mathbf{U}\mathbf{Z}^{(k)}\mathbf{U}^{*}, \qquad (4)$$

where $k$ – block index, and if no compression procedure was applied $\tilde{\mathbf{Y}}^{(k)} = \mathbf{Y}^{(k)}$.

In fig. 1 shows the original raster monochrome square image "barbara.png" (512×512 pixels) and the result of applying the above-described block discrete cosine transform (8×8 pixels) procedure.

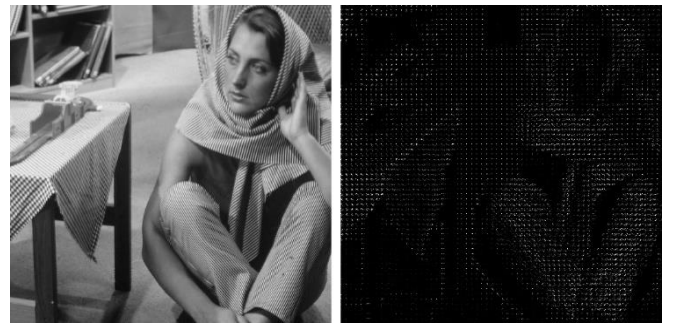

Fig. 1. Original image (left), block DCT (right).

After applying the block DCT (Fig. 1), the quantization operation is applied to each received block, which is described by the expression

$$\tilde{\mathbf{Z}}^{(k)}_{i,j} = \text{round}\left(\dfrac{\mathbf{Z}^{(k)}_{i,j}}{R \cdot \mathbf{Q}_{i,j}}\right), \qquad (5)$$

$$i = 0, ..., M-1, \; j = 0, ..., M-1.$$

where $\mathbf{Z}^{(k)}$ – $k$-th block of the matrix of elements of the spectrum of the image of dimension (8×8), $\mathbf{Q}$ – quantization matrix, $M$ – block dimension (in this case $M = 8$), and $R \in \mathbb{N}^+$ – natural quantization factor by default equals to 1.

The $R$ quantization factor in the formula (5) controls the compression ratio – the higher the value, the stronger the compression. In turn, the quantization matrix $\mathbf{Q}$ (as specified in the original JPEG standard) is defined as follows

$$\mathbf{Q} = \begin{bmatrix} 16 & 11 & 10 & 16 & 24 & 40 & 51 & 61 \\ 12 & 12 & 14 & 19 & 26 & 58 & 60 & 55 \\ 14 & 13 & 16 & 24 & 40 & 57 & 69 & 56 \\ 14 & 17 & 22 & 29 & 51 & 87 & 80 & 62 \\ 18 & 22 & 37 & 56 & 68 & 109 & 103 & 77 \\ 24 & 35 & 55 & 64 & 81 & 104 & 113 & 92 \\ 49 & 64 & 78 & 87 & 103 & 121 & 120 & 101 \\ 72 & 92 & 95 & 98 & 112 & 100 & 103 & 99 \end{bmatrix}.$$

Further, in order to "filter out" the obtained zeros and efficiently encode the resulting image, the so-called zig-zag transform is applied, and then entropy coding (for example, Huffman coding) is used to convert the sequence into bits.

In turn, when restoring an image, all the transformations described above are performed in the reverse order. In addition, instead of the quantization procedure (5), the inverse procedure is used, which is described by the following expression

$$\mathbf{Z}_{i,j}^{(k)} = \tilde{\mathbf{Z}}_{i,j}^{(k)} \left( \mathbf{Q}_{i,j} \cdot R \right), \quad (6)$$
$$i = 0, ..., M-1, \, j = 0, ..., M-1,$$

and the corresponding backward transform is applied instead of the forward block discrete cosine transform.

### III. WEYL-HEISENBERG TRANSFORM

According to early works [10], the unitary complex matrix of the Weyl-Heisenberg orthogonal basis of dimension ($N \times N$) is defined by the expression

$$\mathbf{U} = \Re\{\mathbf{U}_R\} + j\Re\{\mathbf{U}_I\}, \quad (7)$$

whose elements are calculated by the equations

$$\mathbf{U}_R(n, lM+k) = g[(n-lM)_N] e^{2\pi j \frac{k}{M}(n-\alpha/2)},$$

$$\mathbf{U}_I(n, lM+k) = jg[(n+\frac{M}{2}-lM)_N] e^{2\pi j \frac{k}{M}(n-\alpha/2)},$$

$$n = 0, ..., N-1, \, k = 0, ..., M-1, \, l = 0, ..., L-1, \, N = LM,$$

where $L$ – number of time shifts, $M$ – number of frequency shifts, $g(\cdot)$ – optimized forming WH-function of dimension $N$, and $\alpha$ – phase parameter.

Let us clarify that in image compression problems, the use of a complex Weyl-Heisenberg matrix is not appropriate, since we are dealing with a real two-dimensional signal [10].

However, we can use the corresponding real version of matrix (7), which is defined by the expression

$$\tilde{\mathbf{U}} = \Re\{\mathbf{U}\} + \Im\{\mathbf{U}\}. \quad (8)$$

Formula (8) can be written in the equivalent form

$$\tilde{\mathbf{U}} = \Re\{\mathbf{U}_R\} + \Re\{\mathbf{U}_I\}. \quad (9)$$

In early works [10] we show that the transformation matrix (9) is an orthogonal and satisfies property (1).

In order to compare the developed discrete orthogonal Weyl-Heisenberg transform with the block discrete cosine transform in the image compression problem, we take the value of the number of frequency shifts $M = 8$ (which actually corresponds to the 8×8 block separation), then the value of the phase parameter $\alpha = M/2 = 4$. We also write down the formula for calculating the optimal value $\sigma = 1/(M^2\beta)$, where $\beta$ is a parameter providing optimal localization of the Gaussian function.

In fig. 2 shows the raster monochrome square image "barbara.png" (512×512 pixels) and its spectrums for block DCT and DWHT (for $\beta = 2$). For the convenience of image comparison, images are cropped to 16×16 pixels.

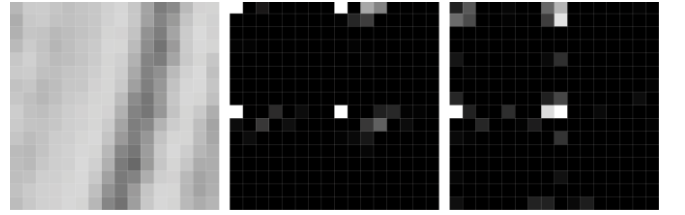

Fig. 2. Original (left), block DCT (center), DWHT (right).

Similar to the JPEG compression algorithm, one can define the quantization matrix $\mathbf{Q}$ for the discrete Weyl-Heisenberg transform

$$\mathbf{Q} = \begin{bmatrix} 8 & 16 & 24 & 32 & 32 & 24 & 16 & 8 \\ 16 & 32 & 48 & 64 & 64 & 48 & 32 & 16 \\ 24 & 48 & 72 & 96 & 96 & 72 & 48 & 24 \\ 32 & 64 & 96 & 128 & 128 & 96 & 64 & 32 \\ 32 & 64 & 96 & 128 & 128 & 96 & 64 & 32 \\ 24 & 48 & 72 & 96 & 96 & 72 & 48 & 24 \\ 16 & 32 & 48 & 64 & 64 & 48 & 32 & 16 \\ 8 & 16 & 24 & 32 & 32 & 24 & 16 & 8 \end{bmatrix}.$$

In this case, the values of the elements of the quantization matrix $\mathbf{Q}$ for the discrete Weyl-Heisenberg transform were selected based on the features of the resulting spectrum.

### IV. EXPERIMENTAL RESULTS

As in the previous work [10], we calculate the Frobenius (Euclidian) norm of the difference between the original image and the one reconstructed after compression to estimate the degree of difference between the original image and its compressed version

$$E = \sqrt{\sum_{i=1}^{m}\sum_{j=1}^{n} \left( \mathbf{Y}_{ij} - \tilde{\mathbf{Y}}_{ij} \right)^2} = \|\mathbf{Y} - \tilde{\mathbf{Y}}\|_F. \quad (10)$$

where $m, n$ – width and height of the image, respectively. The formula (10) will be the main indicator of the difference between the restored image and the original one, and therefore it will be a criterion for quality losses.

Peak signal-to-noise ratio (PSNR) is often used to measure the level of distortion in image compression. Since most signals have a wide dynamic range, *PSNR* is usually measured on a logarithmic scale in decibels and is defined as

$$MSE = \frac{1}{mn} \sum_{i=0}^{m-1} \sum_{j=0}^{n-1} \left| \mathbf{Y}_{ij} - \tilde{\mathbf{Y}}_{ij} \right|^2,$$

$$PSNR = 10\log_{10}\left(\frac{MAX_\mathbf{Y}^2}{MSE}\right), \quad (11)$$

where *m*, *n* – width and height of the image.

We introduce the compress ratio *K*, which means the ratio of the number of zero elements of the image spectrum to the total number of image elements (in percent)

$$K = \left(N_Z / N_T\right) \cdot 100\%. \quad (12)$$

where $N_Z$ – number of zero elements of the image spectrum, $N_T$ – total number of image spectrum elements ($N_T = mn$).

To compare DCT (according to the JPEG specification) and DWHT (at β = 2) in the problem of image compression, modeling was carried out for different values of the quantization coefficient *R* = 1, 4, 8 and the values of indicators described by formulas (10-12). Figures 3-4 show, as an example, the results of compressing the same image "barbara.png" (512×512 pixels) at *R* = 1, 4, 8 using DCT (JPEG) and DWHT, and Tables 1-2 the corresponding numerical results are presented.

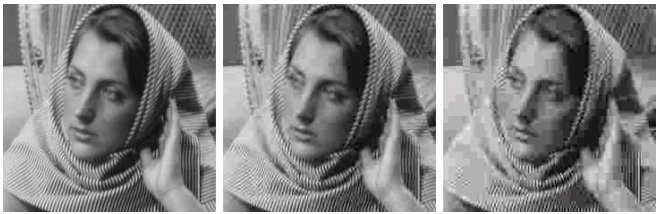

Fig. 3. Compression using DCT (JPEG) for *R* = 1, 4, 8.

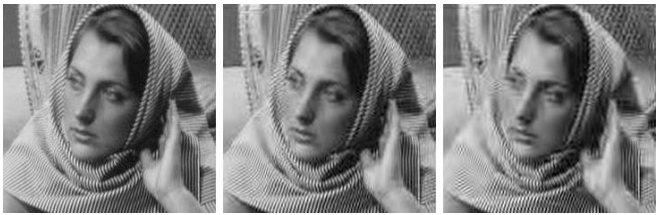

Fig. 4. Compression using DWHT (β = 2) for *R* = 1, 4, 8.

TABLE I. COMPRESSION USING DCT (JPEG)

| Quantization coefficient, *R* | 1 | 4 | 8 |
|---|---|---|---|
| Compression coefficient, *K* (%) | 83.73 | 93.82 | 96.63 |
| PSNR (dB) | 32.53 | 26.26 | 23.92 |
| Quality losses, *E* | 438.58 | 980.48 | 1399.16 |

TABLE II. COMPRESSION USING DWHT (β = 2)

| Quantization coefficient, *R* | 1 | 4 | 8 |
|---|---|---|---|
| Compression coefficient, *K* (%) | 85.68 | 94.68 | 96.87 |
| PSNR (dB) | 33.87 | 27.65 | 25.29 |
| Quality losses, *E* | 336.66 | 776.45 | 1106.21 |

In the case of the Weyl-Heisenberg transform (DWHT), we have the lowest quality loss of the compressed image. This can be seen from the comparative analysis of visual and numerical characteristics. According to the simulation results, DWHT provides an average of 20-25% preservation of image quality compared to DCT (JPEG) at a higher compression ratio *K*.

V. CONCLUSION

Based on the simulation results, it can be concluded that the presented new approach to compression of raster images, based on the real DWHT, provides a high compression ratio and greater preservation of image quality during restoration in comparison with the well-known JPEG compression standard. Consequently, the use of the DWHT turns out to be a very effective tool in the problem of compressing and storing raster images. This phenomenon is explained by the fact that a discrete image is essentially a non-stationary two-dimensional random process. The features of this process can be fully studied only in the time-frequency domain, having previously broken it into fragments. It is this fact that is critical for filtering out non-essential spectral components and, as a result, effective image compression.